\documentclass[aps,pra,reprint,superscriptaddress,nofootinbib,citeautoscript]{revtex4-2}
\setcitestyle{super}

\usepackage{microtype}
\usepackage[print-unity-mantissa=false, separate-uncertainty=true]{siunitx}
\usepackage{physics}
\usepackage{booktabs}

\usepackage{pgfplots}
\pgfplotsset{compat=newest}
\usepgfplotslibrary{external, fillbetween, colorbrewer, groupplots, units, polar}
\usetikzlibrary{external, shadows, arrows.meta, decorations.markings, fadings, patterns}
\pgfplotsset{compat=newest,
	/pgf/number format/1000 sep={\,},
	tick label style={font=\footnotesize},
	label style={font=\small},
	legend style={font=\footnotesize, draw=none, fill=none},
	height=6cm,
	width=\columnwidth,
	major grid style={lightgray, ultra thin},
	legend cell align={left},
	legend image code/.code={\draw[mark repeat=2,mark phase=2] plot coordinates {(0cm,0cm) (0.12cm,0cm) (0.24cm,0cm)};},
	filter discard warning=false,
	xlight/.style={Paired-E, semithick},
	xdark/.style={Paired-F, semithick},
	ylight/.style={Paired-C, semithick},
	ydark/.style={Paired-D, semithick},
}

\newif\ifuseprecompiledfigures
\useprecompiledfigurestrue 

\begin{document}

\graphicspath{{./figures}}
\title{Experimental determination of effective light transport properties in fully anisotropic media}

\author{Ernesto Pini}
\email{ernesto.pini@unifi.it}
\affiliation{Department of Physics and Astronomy, Universit\`{a} di Firenze, Sesto Fiorentino, Italy}
\affiliation{European Laboratory for Non-linear Spectroscopy (LENS), Sesto Fiorentino, Italy}

\author{Peter Nagli\v{c}}
\affiliation{University of Ljubljana, Laboratory of Imaging Technologies, Faculty of Electrical Engineering, Ljubljana, Slovenia}

\author{Miran B\"urmen}
\affiliation{University of Ljubljana, Laboratory of Imaging Technologies, Faculty of Electrical Engineering, Ljubljana, Slovenia}

\author{Alexander Gatto}
\affiliation{Sony Europe B.V., Stuttgart Technology Center, Stuttgart, Germany\\}

\author{Henrik Sch\"afer}
\affiliation{Sony Europe B.V., Stuttgart Technology Center, Stuttgart, Germany\\}

\author{Diederik S. Wiersma}
\affiliation{Department of Physics and Astronomy, Universit\`{a} di Firenze, Sesto Fiorentino, Italy}
\affiliation{European Laboratory for Non-linear Spectroscopy (LENS), Sesto Fiorentino, Italy}
\affiliation{Istituto Nazionale di Ricerca Metrologica (INRiM), Turin, Italy}

\author{Lorenzo Pattelli}
\email{l.pattelli@inrim.it}
\affiliation{Istituto Nazionale di Ricerca Metrologica (INRiM), Turin, Italy}
\affiliation{European Laboratory for Non-linear Spectroscopy (LENS), Sesto Fiorentino, Italy}

\date{\today}

\begin{abstract}
Structurally anisotropic materials are ubiquitous in several application fields, yet their accurate optical characterization remains challenging due to the lack of general models linking their scattering coefficients to the macroscopic transport observables, and the need to combine multiple measurements to retrieve their direction-dependent values. Here, we present an improved method for the experimental determination of light transport tensor coefficients from the diffusive rates measured along all three directions, based on transient transmittance measurements and a generalized Monte Carlo model. We apply our method to the characterization of light transport properties in two common anisotropic materials -- polytetrafluoroethylene (PTFE) tape and paper -- highlighting the magnitude of systematic deviations that are typically incurred when neglecting anisotropy.
\end{abstract}

\maketitle

\section{Introduction}

Turbid media are encountered in several applied and fundamental research fields, where determining the light transport properties of these materials is important to study their microscopic structure and composition\cite{martelli2022light}.
Among several experimental techniques, time-domain methods are particularly useful as they allow to isolate contributions dominating at different time scales, ranging from early ballistic light to the late multiple scattering regime\cite{mazzamuto2016deducing, svensson2013exploiting}.
Despite the importance of an accurate determination of these properties, however, the experimental retrieval of scattering parameters is typically performed under the assumption that light transport is isotropic -- even when the material under study is visibly anisotropic\cite{strudley2013mesoscopic, burresi2014bright, konagaya2016optical, yang2022cellulose, moradi2023monte}. Ignoring the presence of anisotropic light transport clearly introduces an error in the determination of the microscopic scattering properties, the magnitude of which is currently not well characterized.

This bias can have important consequences for the characterization of tissues, fibrous materials, and strongly scattering media -- which often exhibit an anisotropic structure or morphology.
In the large majority of cases, the scattering strength of these materials has been characterized using a single (scalar) transport mean free path, which however can be quite far from the value of the actual components of their scattering tensor, and even from their direction-averaged value.
Additionally, the experimental methods that are used to infer a single mean free path value for an anisotropic medium are inevitably more sensitive to the scattering strength along a certain direction.
Thus, whenever an anisotropic medium is characterized via a single scalar mean free path, the resulting value may be more or less indicative of the scattering properties in the plane\cite{kienle2007determination, muskens2009large, cortese2015anisotropic, konagaya2016optical, moffa2018biomineral} or those along the perpendicular direction\cite{burresi2014bright, toivonen2018anomalous, zou2019biomimetic, yang2022cellulose, strudley2013mesoscopic}, rather than being representative of the whole medium.

For isotropic materials, a simple relation exists linking the transport mean free path $\ell^*$ to the (experimentally observable) diffusive constant $D = v\ell^*/3$, with $v$ as the energy velocity in the medium.
At the microscopic level, this isotropic transport mean free path is in turn related to the scattering mean free path $\ell_\text{s}$ (defined as the inverse of the scattering coefficient $\ell_\text{s} = 1/\mu_\text{s}$) via a ``similarity relation'' $\ell^* = \ell_\text{s}/(1 - g)$ involving the cosine of the average polar scattering angle $g$.
Therefore, for an isotropic medium, the relation between the macroscopic observable diffusion rate and the microscopic scattering property can be written as:
\begin{equation} \label{eq:D}
	D = \frac{1}{3} v \ell^* =  \frac{1}{3} \frac{v}{\mu_\text{s}'} = \frac{1}{3} \frac{v}{\mu_\text{s} (1 - g)},
\end{equation}
with $\mu_\text{s}' = \mu_\text{s} (1 - g)$ defined as the reduced scattering coefficient.

For anisotropic materials, however, this relation is not strictly valid any more \cite{alerstam2014anisotropic, pini2024diffusion}.
Additionally, all parameters of interest can in principle become $3 \times 3$ tensor quantities, including $\vb*{D}$, $\vb*{\mu_\text{s}}$, $\vb*{g}$ and $\vb*{v}$.
Assuming for simplicity that all tensors can be diagonalized in a common reference frame, one can still define the diagonal components of the scattering tensor as $\mu_{\text{s}, i} = 1/\ell_{\text{s}, i}$ and diffusive rate tensor as $D_i = v_i^{\vphantom{*}} \ell_i^* /3$, with $i \in \{x, y, z\}$.
However, for fully anisotropic materials ($\mu_{\text{s}, x}\neq\mu_{\text{s}, y}\neq\mu_{\text{s}, z}$), these quantities are no longer related to each other via simple analytical relationships, and the diffusive rate along a certain direction will not depend solely on the scattering properties along the same direction.

To elucidate this gap, multiple competing descriptions of anisotropic transport have been proposed in the past years \cite{heino2003anisotropic, kienle2007anisotropic, alerstam2014anisotropic, vasques2014non, han2020transport}, which however do not provide a general link between microscopic scattering parameters and the corresponding diffusion observables.
For this reason, a trade-off must be introduced that allows to properly account for transport anisotropy, while keeping the number of tensor quantities involved in the inverse problem to a minimum to allow their practical determination.

To improve on the current situation and avoid the use of oversimplified isotropic assumptions, we introduce a microscopic effective transport tensor and present a method to determine its components for scattering media with full anisotropy in all three spatial directions.
To this purpose, we assume a tensor scattering coefficient $\vb*{\mu_\text{s}}$, while keeping the asymmetry factor $g$, absorption coefficient $\mu_\text{a}$ and refractive index $n$ as scalar quantities.
Effective transport mean free path values $\tilde{\ell}_i^*$ are then still defined in analogy with the standard similarity relation via the scalar $g$ value, as needed for materials characterized by a limited optical thickness or a strong forward scattering:
\begin{equation}
	\tilde{\ell}_i^* = \frac{1}{\tilde{\mu}_{\text{s}, i}^\prime} = \frac{1}{\mu_{\text{s}, i}(1 - g)}.
\end{equation}
Defining an effective transport mean free path is of practical utility when dealing with turbid samples, as it allows to account for the transient effects due to scattering asymmetry while leaving the overall diffusion tensor $\vb*{D}$ almost unaffected for small variations of $g$.
This approach captures the key features of anisotropic transport, such as the expected discrepancy between the observed diffusive rates (expressed in units of length as transport mean free paths $\ell_i^*$) and their effective microscopic counterparts $\tilde{\ell}_i^*$, which becomes more significant with increasing structural anisotropy.
More fundamentally, it allows to restore a link between a random walk description of transport and the resulting diffusive rates, by assigning different identities to these two parameters.

We apply this approach to experimental measurements obtained using an optical gating technique \cite{pattelli2016spatio, pattelli2022experimental, pini2024diffusion, pini2023breakdown}, and analyze the results using a newly developed and open-source Monte Carlo package named PyXOpto \cite{naglic2021pyxopto} which is capable of handling tensor scattering coefficients.

The paper is organized as follows: in Section \ref{sec:anisdeg}, we start by commenting on the typical degree of apparent degeneracy that can be expected between isotropic and anisotropic transport, by showing a few representative examples of time-domain transmittance through a slab geometry.
Resorting to an isotropic model or theory in this configuration can lead to significant systematic errors even when time-resolved measurements are limited to the retrieval of scattering properties along the depth direction.
Experimental and numerical methods are described in Section \ref{sec:methods}.
Results are presented in Section \ref{sec:results} relative to the determination of the transport properties of two common laboratory materials, providing a quantitative illustration of the typical errors incurred when disregarding anisotropic light transport.
Finally, conclusions and perspectives are drawn in Section \ref{sec:conclusions}.

\section{Decay rate degeneracy} \label{sec:anisdeg}

In many experimental configurations of interest, a sample is illuminated from the perpendicular direction to its interface, and the decay rate of the intensity transmitted or reflected from the sample interface is studied.
In the diffusive regime, this decay rate is determined by the diffusive coefficient along the perpendicular axis.
At the microscopic level, however, this coefficient depends on the scattering properties along all directions\cite{alerstam2014anisotropic, pini2024diffusion}, meaning that it is not possible to reliably infer the scattering coefficient along the perpendicular direction from a time-domain measurement alone.

\begin{figure*}
	\centering
	\ifuseprecompiledfigures
		\includegraphics{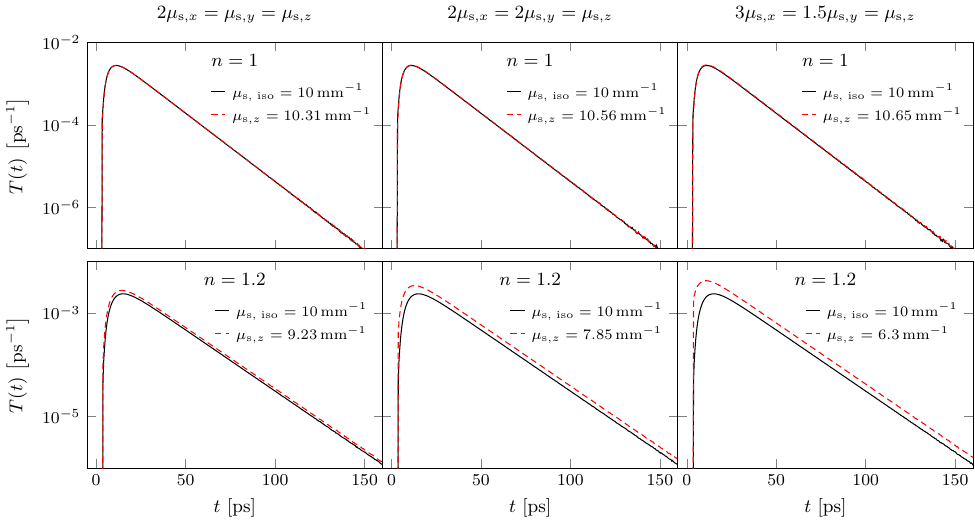}
	\else
		\input{figures/Degeneracy.tikz}
	\fi
	\caption{Time-resolved transmittance $T(t)$ through a non-absorbing \SI{1}{\milli\meter}-thick scattering slab for two refractive index contrast values $n = 1, 1.2$. Solid curves represent isotropic simulations with $\mu_{\text{s, iso}} = \SI{10}{\per\milli\meter}$. For each case a pre-defined ratio between scattering coefficients along different axes is imposed, as specified in the title above each column. Anisotropic configurations exhibiting the same decay rate as the isotropic case are shown, leading to values of $\mu_{\text{s},z} \neq \mu_{\text{s, iso}}$. All simulations are performed setting $g = 0$.}
	\label{fig:degeneracy}
\end{figure*}

To exemplify this issue, we consider a set of Monte Carlo simulations showing that several different anisotropic configurations can give rise to transmittance decay rates that are degenerate with the isotropic case, despite using different scattering coefficients along the illumination direction $z$.
Taking a non-absorbing \SI{1}{\milli\meter}-thick slab with isotropic scattering $\mu_{\text{s, iso}} = \SI{10}{\per\milli\meter}$ and refractive indexes of $n = 1$ or $n = 1.2$, one can find alternative configurations with pre-defined degrees of anisotropy (encompassing both cases of partial or full 3D anisotropy) which give rise to the same transmittance decay rate (Figure \ref{fig:degeneracy}).

Despite the different scattering coefficients, the resulting curves are nearly identical except for a constant scaling factor and the initial transient.
Experimentally, these differences are not decisive to tell apart isotropic from anisotropic samples, since time-resolved curves are typically measured with arbitrary units, and the details of the early transient are often concealed by the finite duration of the illumination pulse.
For the representative configurations shown here, deviations between the isotropic scattering coefficient $\mu_{\text{s}}$ and the longitudinal one $\mu_{\text{s}, z}$ range from a few percent in the index-matched cases, up to \SI{37}{\percent} in the presence of full 3D anisotropy and moderate refractive index contrast.
It should be further noted that if the absorption coefficient is also unknown and left as a free parameter, almost perfect degeneracy among all curves can be typically obtained at all times, leading to even larger discrepancies between the true scattering coefficient and its corresponding isotropic value.

These examples show that, for any practical purposes, time-resolved curves alone have limited informative value for the study of structurally anisotropic media, and need to be complemented with additional independent measurements.

\section{Methods} \label{sec:methods}

\subsection{Experimental setup}
\begin{figure*}
	\centering
	\includegraphics[width=\linewidth]{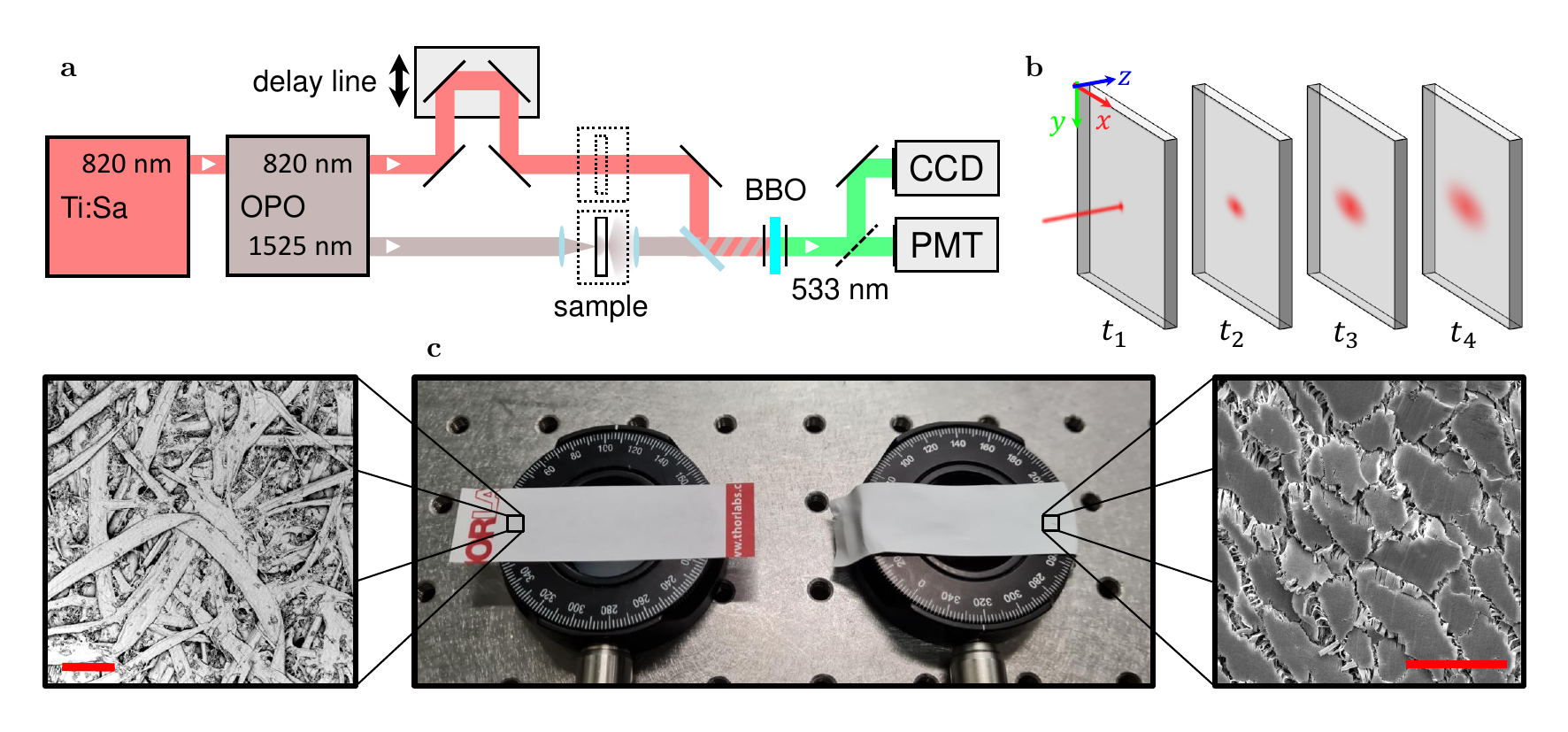}
	\caption{(a) Sketch of the experimental setup for time-resolved measurements when using a \SI{1525}{\nano\meter} probing wavelength: Ti:SA, Titanium Sapphire \si{\femto\second} laser; OPO, optical parametric oscillator; BBO, $\beta$-barium borate non-linear crystal; CCD, charge-coupled device camera; PMT, photomultiplier tube. (b) Schematics of the transient imaging measurement on a slab sample. The instantaneous spatial distribution of transmitted intensity is imaged at different time delays. (c) Photograph of the paper post-it note and PTFE tape strip with corresponding scanning electron microscope images of the two samples (insets). Scale bars correspond to \SI{100}{\micro\meter} and \SI{30}{\micro\meter}, respectively.}
	\label{fig:setup}
\end{figure*}

We approach the study of anisotropic materials using an experimental setup based on the transient imaging optical gating method, which is capable of providing both spatial and time-domain information on the evolution of intensity profiles \cite{pattelli2016spatio, pattelli2022experimental}.
A Ti:Sa laser source and a parametric oscillator generate two near-infrared synchronous trains of pulses at \SI{820}{\nano\meter} and \SI{1525}{\nano\meter}, with a \SI{80}{\mega\hertz} repetition rate and a typical duration of about \SI{150}{\femto\second} (Fig.\ \ref{fig:setup}a).
In the experiment, a motorized stage is used to adjust the relative delay between the two laser beams, which can either serve as the probe or the gate arms of the setup.
The probe beam is focused on the sample to generate a diffused transmittance signal, which is then recombined with the (expanded) gate beam onto a $\beta$-barium borate (BBO) crystal.
A sum-frequency signal is finally generated at a wavelength of \SI{533}{\nano\meter}, proportional to the cross-correlation amplitude between the diffuse transmittance and the gate pulse at a given delay.
The resulting up-converted time-resolved signal can be then either integrated by a photomultiplied detector, or resolved spatially with a CCD camera recording the transient intensity profiles transmitted through the scattering medium.
Thanks to the spatially uniform upconversion efficiency guaranteed by the expanded and collinear gate beam, the generated images provide quantitative information on the transverse propagation light at the exit surface of the sample, from which the diffusive rates along different directions can be estimated directly.
All measurements are averaged over multiple sample positions in an area of $\sim \SI{4}{\milli\meter\squared}$, to reduce fluctuations due to small local inhomogeneities and speckle patterns.

\subsection{Data analysis and Monte Carlo simulations}
Intensity profiles are recorded at different time delays, as exemplified in Fig.\ \ref{fig:setup}b.
The profiles are rotated for convenience to align the intrinsic sample anisotropy axes to the laboratory reference frame, and then fitted with bi-variate Gaussian distributions to retrieve the instantaneous Mean Square Displacement (MSD) along the $y$ (vertical) and $x$ (horizontal) axes.
The spread rate of the intensity profiles provides direct access to the $x$ and $y$ diffusion rates, independent of any absorption contribution \cite{mazzamuto2016deducing, pini2024diffusion}.
Thus, performing a linear regression on the MSD evolution after the initial transient allows to retrieve the diffusive constants $D_x$, $D_y$, which are proportional to the corresponding slopes.
Information on the diffusive rate along $z$ (and the absorption coefficient $\mu_\text{a}$) is finally extracted by integrating the total transmitted intensity in each frame, or using a photo-multiplier tube for convenience.

In order to retrieve the microscopic scattering coefficients associated to the observed diffusive rates, anisotropic Monte Carlo simulations are performed by varying the scattering coefficient tensor components until the experimental MSD growth and the time-resolved transmittance curves are simultaneously reproduced with a single simulation.
To this purpose, we used the PyXOpto open-source implementation of the Monte Carlo method which is capable of handling tensor scattering properties \cite{naglic2021pyxopto}.
In the fitting procedure, the three components of the scattering tensor $\mu_{\text{s}, x}$, $\mu_{\text{s}, y}$ and $\mu_{\text{s}, z}$ are used as independent free parameters, plus a scalar asymmetry factor $g$ which is applied to all three axes.

\section{Experimental results} \label{sec:results}

We characterize light transport experimentally in two common materials: PTFE (Teflon) tape and paper, both of which are generally known for their highly anisotropic structure and vanishing absorption at visible and near-infrared wavelengths.
Notably, paper and PTFE tape are also commonly indicated as reference diffusers in the evaluation protocols for the temporal performance of near-infrared spectroscopy applications, based on the assumption that they contribute negligibly to the temporal dispersion of the illumination pulse.\cite{wabnitz2014performance}
However, to the best of our knowledge, the transport mean free path in these materials is not known precisely, as previous time-resolved measurements were unable to appreciate their multiple scattering regime\cite{carlsson1995time}.

In the following, we perform time-domain transmittance measurements through these samples, and analyze the results using both isotropic and anisotropic Monte Carlo simulations, which allows us to determine the magnitude of the error introduced by not accounting properly for the structural anisotropy of the materials.

\subsection{PTFE tape}
The first sample we consider is a strip of Teflon tape commonly used for sealing pipe threads.
Teflon tape is made by polytetrafluoroethylene (PTFE) fibers exhibiting a preferential alignment, which endow it with a marked structural anisotropy.
This anisotropy affects also light transport properties, as confirmed by previous reports on the observation of direction-dependent light diffusion in this material\cite{johnson2009optical, badon2016spatio}.

We measure the transmittance through a free-standing strip of tape at a probe wavelength of \SI{1525}{\nano\meter}.
The tape sample is attached on a rotating mount to allow its alignment with the laboratory reference frame (Figure \ref{fig:setup}c) and verify that the elongated transmittance profiles rotate integrally with the sample.
The analysis reveals the presence of a direction of faster diffusion at an angle of $\ang{\sim 45}$ with respect to the direction of the tape strip.
This angle corresponds to the direction perpendicular to the alignment of the PTFE fibers, as determined by scanning electron microscopy images, in agreement with previous reports \cite{johnson2009optical}. 
Closer inspection of the electron microscopy picture shows that the polymer fibers have a flattened shape along the main plane of the tape, and are further fused together to form planar flakes on its external surface.
This morphology suggests that different scattering coefficients could be found along each direction, resulting in a fully anisotropic sample.

\begin{figure*}
	\centering
	\ifuseprecompiledfigures
		\includegraphics{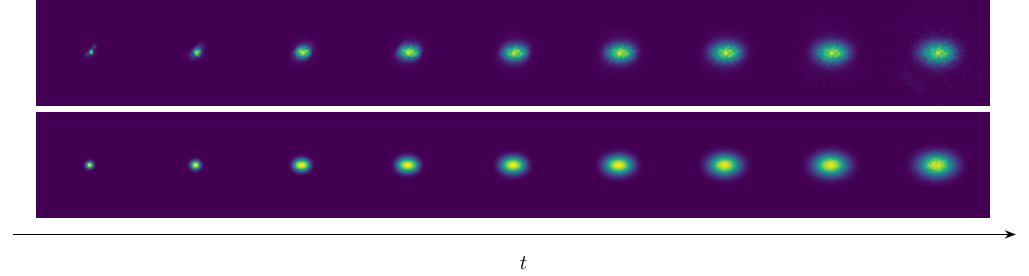}
	\else
		\input{figures/ImagingTeflon.tikz}
	\fi
	\caption{Selection of transient imaging measurements (top) and corresponding best-fit MC simulation (bottom) of the transmitted intensity distribution recorded at different times for the PTFE tape sample probed at \SI{1525}{\nano\meter}. Each frame shows an area of $7.4 \times \SI{7.4}{\milli\meter}$ and is normalized to its maximum intensity value. The time delay between consecutive frames is $\Delta t = \SI{1.33}{\pico\second}$.}
	\label{fig:imagingTeflon}
\end{figure*}
\begin{figure*}
	\centering
	\ifuseprecompiledfigures
		\includegraphics{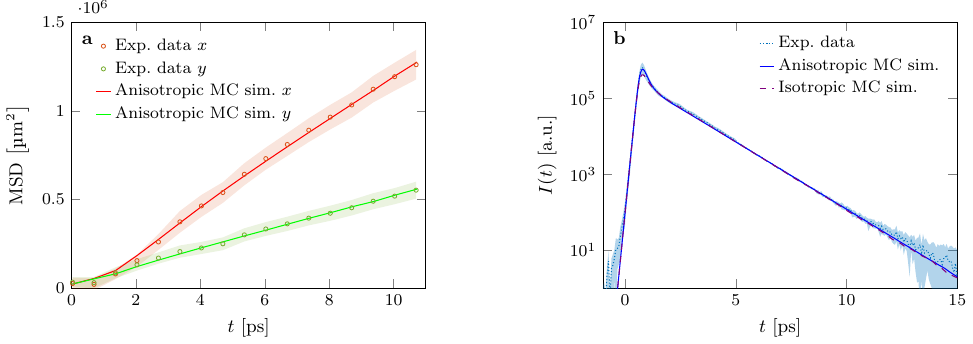}
	\else
		\input{figures/Teflon.tikz}
	\fi
	\caption{(a) Mean square displacement and (b) time-resolved transmitted intensity of PTFE tape. Solid lines represent the best-fit anisotropic MC simulation, obtained with: $\mu_{\text{s}, x} = \SI{48}{\per\milli\meter}$, $\mu_{\text{s}, y} = \SI{105}{\per\milli\meter}$, $\mu_{\text{s}, z} = \SI{77}{\per\milli\meter}$, $g = 0.9$, $\mu_\text{a} = 0$. The dashed purple curve shows a fit performed with an isotropic MC simulation with $\mu_\text{s, iso} = \SI{95}{\per\milli\meter}$. Shaded areas represent (a) the compounded uncertainty from the pixel-to-\si{\micro\meter} conversion and the bi-variate Gaussian fit, and (b) $1\sigma$ of the average between 5 different sample positions, respectively.}
	\label{fig:resultsTeflon}
\end{figure*}

The main set of transient imaging measurements was performed in a fixed orientation configuration (Fig.\ \ref{fig:imagingTeflon}), rotated so that the faster axis of diffusion coincides with the $x$ axis in the laboratory reference frame.
An average refractive index of $n = \num{1.05}$ was evaluated for the slab using the Bruggeman mixing rule \cite{lakhtakia1997bruggeman} based on the PTFE/air volume fraction estimated by a weighting method.
Sample thickness was set to $L = \SI{200}{\micro\meter}$, as declared by the tape manufacturer.
The diffusion coefficients can be derived directly with a linear regression on the MSD evolution after the initial transient for $x$ and $y$ (Fig.\ \ref{fig:resultsTeflon}a), returning values of $D_x = \SI{289(7)e2}{\micro\meter\squared\per\pico\second}$ and $D_y = \SI{117(2)e2}{\micro\meter\squared\per\pico\second}$.

The resulting best-fit MC simulation is in excellent agreement with both the MSD evolution and the time-resolve transmitted intensity, reproducing also finer experimental features such as the slight slope change of the fast-diffusing axis after $t \sim \SI{5}{\pico\second}$ (Fig.\ \ref{fig:resultsTeflon}a) or the more prominent transmittance peak at $t \sim \SI{1}{\pico\second}$ (Fig.\ \ref{fig:resultsTeflon}b).

The corresponding effective transport mean free paths values are $\tilde{\ell}_x^* = \SI{210(20)}{\micro\meter}$, $\tilde{\ell}_y^* = \SI{95(7)}{\micro\meter}$, $\tilde{\ell}_z^* = \SI{130(14)}{\micro\meter}$, which remain basically unaltered when forcing different values of $g$ around its fitted value of \num{0.9}.

It is interesting to note that the ratio between the diffusive constants $D_x/D_y = \num{2.47}$ is appreciably different from the ratio between the effective transport mean free paths $\tilde{\ell}_x^*/\tilde{\ell}_y^* = \num{2.21}$, which confirms the different roles taken by these two parameters, and shows that a direct measurement of the diffusion dynamics does not even provide relative information on the microscopic degree of anisotropy due to the non trivial mixing of the scattering coefficients along all directions.

An attempt to fit the time-resolved experimental data is also performed assuming an isotropic Monte Carlo model, using the previously determined values for $n$, $\mu_{\text{a}}$ and $g$, with $\mu_\text{s, iso}$ as the only free parameter.
The result shows a comparable agreement (Fig.\ \ref{fig:resultsTeflon}b), returning however a value that differs by \SI{24}{\percent} from the actual scattering coefficient along the $z$ axis, as summarized in Tab.\ \ref{tab:results}.

\subsection{Paper}

\begin{figure*}
	\centering
	\ifuseprecompiledfigures
		\includegraphics{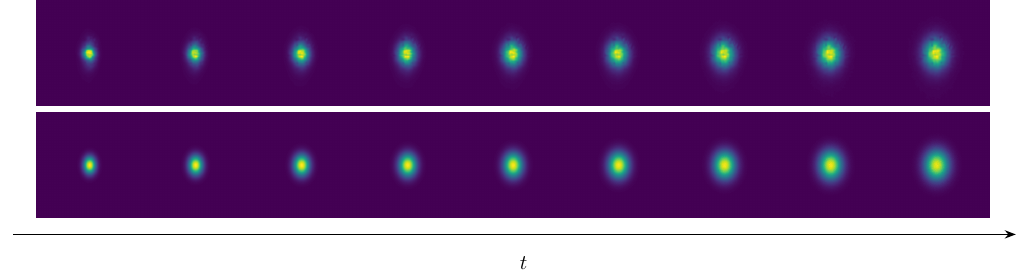}
	\else
		\input{figures/ImagingPaper.tikz}
	\fi
	\caption{Selection of transient imaging measurements (top) and corresponding best-fit MC simulation (bottom) of the transmitted intensity distribution recorded at different times for the post-it note performed at \SI{820}{\nano\meter} probing wavelength. Each frame shows an area of $2.7 \times \SI{2.7}{\milli\meter}$ and is normalized to its maximum value. The time delay between consecutive frames is $\Delta t = \SI{1}{\pico\second}$.}
\label{fig:imagingPaper}
\end{figure*}

The second anisotropic sample that we study is a piece of common paper, which is a material known for its structural anisotropy arising form the preferential alignment of its cellulose fibers imparted during its industrial fabrication process \cite{linder2012anisotropic, linder2013lateral, linder2014light}.
A post-it note was tested in a transmission configuration at \SI{820}{\nano\meter} probing wavelength, aligning the fast diffusion direction along the $y$ axis (Fig.\ \ref{fig:imagingPaper}).

\begin{figure*}
	\centering
	\ifuseprecompiledfigures
		\includegraphics[width=\linewidth]{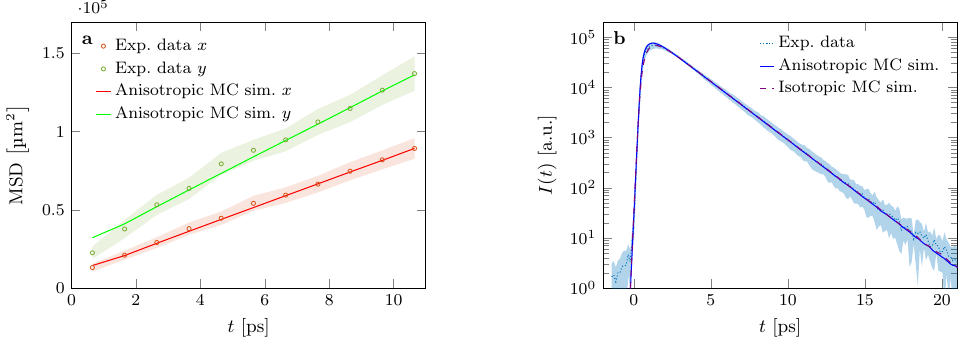}
	\else
		\input{figures/Paper.tikz}
	\fi
	\caption{(a) Mean square displacement and (b) time-resolved transmitted intensity of paper. Solid lines represent the best-fit anisotropic MC simulation, obtained with: $\mu_{\text{s}, x} = \SI{223}{\per\milli\meter}$, $\mu_{\text{s}, y} = \SI{148}{\per\milli\meter}$, $\mu_{\text{s}, z} = \SI{343}{\per\milli\meter}$, $g = 0.8$, $\mu_\text{a} = \SI{0.015}{\per\milli\meter}$. The dashed purple curve shows a fit performed with an isotropic MC simulation with $\mu_\text{s, iso} = \SI{455}{\per\milli\meter}$. Shaded areas represent (a) the compounded uncertainty from the pixel-to-\si{\micro\meter} conversion and the bi-variate Gaussian fit, and (b) $1\sigma$ of the average between 6 different sample positions, respectively.}
	\label{fig:resultsPaper}
\end{figure*}

Sample thickness was determined by directly measuring a stack of 50 post-it notes using a precision micrometer, yielding a value of $L = \SI{100(5)}{\micro\meter}$, while its effective refractive index was evaluated at $n = \num{1.25}$ using the same method applied for the PTFE tape.
In contrast with the previous case, a small but measurable absorption coefficient of $\mu_{\text{a}} = \SI{0.015}{\per\milli\meter}$ was independently estimated from a time-resolved measurement of light diffusely reflected by a thick stack of post-it notes, which we applied to the subsequent analysis together with a value of $g = \num{0.8}$, in agreement with previous estimates in the literature\cite{modric2009monte, coppel2011lateral}.
The diffusive coefficients measured from the MSD growth rates (Fig.\ \ref{fig:resultsPaper}a) are $D_x = \SI{18.8(1.1)e2}{\micro\meter\squared\per\pico\second}$ and $D_y = \SI{26(2)e2}{\micro\meter\squared\per\pico\second}$, where the larger uncertainty compared to the case of the PTFE tape is determined by the larger inherent inhomogeneity of paper, which exhibits slight density and thickness fluctuations at different positions.

As in the previous case, the resulting best-fit anisotropic model shows a good agreement with both the MSD evolution and the time-resolved measurement (Fig.\ \ref{fig:resultsPaper}b), excluding an initial transient affected by the local inhomogeneities at the illumination spot.
The fit returned effective scattering mean free paths values of $\tilde{\ell}_x^* = \SI{22.4(1.8)}{\micro\meter}$, $\tilde{\ell}_y^* = \SI{34(3)}{\micro\meter}$ and $\tilde{\ell}_z^* = \SI{14.6(1.7)}{\micro\meter}$.
A ratio of $D_y/D_x = 1.38$ is found between the diffusive coefficients, to be compared with a value of $\tilde{\ell}_y^*/\tilde{\ell}_x^* = 1.52$ for the corresponding effective mean free paths.

The best fit obtained using an isotropic MC model is also shown in Figure \ref{fig:imagingTeflon}b, showing also in this case very good agreement at the cost of a \SI{36}{\percent} systematic error in the retrieved scattering coefficient.
All fitted parameters are summarized in Tab.\ \ref{tab:results}.

\begin{table} [h]
	\caption{Effective transport mean free paths $\tilde{\ell}_i^*$ along different directions retrieved with anisotropic MC simulations for PTFE tape and paper. The transport mean free paths retrieved with isotropic MC modeling ($\ell^*_\text{iso}$) are also reported to show the degree of error introduced by neglecting the presence of anisotropy.}
	\label{tab:results}
	\renewcommand{\arraystretch}{2}%
		\begin{tabular}{cSSSS}
			\toprule
			Sample	& {$\tilde{\ell}_x^*$ [\si{\micro\meter}]}	& {$\tilde{\ell}_y^*$ [\si{\micro\meter}]}	& {$\tilde{\ell}_z^*$ [\si{\micro\meter}]}	& {$\ell^*_\text{iso}$ [\si{\micro\meter}]}\\
			\midrule
			PTFE	& 210(20) 									& 95(7)										& 130(14)									& 105(13) \\
			Paper	& 22.4(1.8)									& 34(3)										& 14.6(1.7)									& 11(1) \\
			\bottomrule
		\end{tabular}
\end{table}

\section{Discussion} \label{sec:conclusions}
In this work, we presented optical gating as a versatile transient imaging method for the study of direction-dependent diffusion of light in structurally anisotropic materials.
The experimental intensity profiles can be directly compared against the output of a general and open-source Monte Carlo description of anisotropic light propagation to retrieve the full transport coefficient tensor of arbitrary anisotropic materials.
We demonstrate our analysis for two common materials exhibiting different diffusive rates along each axis, showing how their time-resolved traces can be easily reproduced with either isotropic or anisotropic Monte Carlo models.
Due to the interplay among all components of the scattering tensor, assuming an isotropic model for an anisotropic material can lead to a substantial error in the determination of the true transport mean free path of a scattering sample, even when the measurement is intended to retrieve only the component along the normal direction to the sample interface.
It is also interesting to note that, in the examples studied here, the transport mean free path retrieved under the (incorrect) isotropic transport assumption is typically shorter than the direction-averaged value, and in the case of paper, even shorter than the shortest component of the corresponding anisotropic mean free path tensor.

Despite the significant advancement offered by new experimental techniques and numerical methods, the complete characterization of structurally anisotropic scattering materials presents still several open challenges.
Significant cross-talk is expected, for instance, when approaching the inverse problem for samples exhibiting anisotropic scattering coefficients combined with birefringence or tensor scattering asymmetry.
This is commonly the case for several common materials, such as biological tissues\cite{wood2007polarized, sun2010determination, tuchin2016polarized, ghassemi2016new} or liquid crystals\cite{wiersma1999time, mertelj2007anisotropic} -- which will likely require an even larger number of independent measurement to constrain the inverse problem associated to the accurate retrieval of their optical properties -- but also for media presenting different orientations of their structural anisotropy at different locations, such as the brain\cite{hebeda1994light, heiskala2005modeling, azimipour2014extraction}, teeth\cite{kienle2002light, zoller2018parallelized} or skin under mechanical deformation\cite{zhang2020effect, ahmed2023investigating}.

Our results highlight the importance of taking into account the presence of structural anisotropy in scattering materials, and the need to investigate all directions rather than just planar or perpendicular propagation, as they all contribute to determining the actual transport mean free path along a direction of interest.
Additionally, we further quantified the potential errors incurred when using an isotropic model for the interpretation of anisotropic transport data, which show a tendency to overestimate the true scattering strength in the case of turbid anisotropic materials.
Notably, this approach has been often used for the characterization of highly scattering materials and their inter-comparison in terms of scattering efficiency along the thickness direction, which calls for a careful reanalysis of previously published results in light of this potential systematic bias.

\section*{Acknowledgments}
This work was partially funded by the European European Union's NextGenerationEU Programme with the I-PHOQS Research Infrastructure [IR0000016, ID D2B8D520, CUP B53C22001750006] ``Integrated infrastructure initiative in Photonic and Quantum Sciences''. E.P acknowledges financial support from Sony Europe B.V.. L.P. further acknowledges the CINECA award under the ISCRA initiative, for the availability of high performance computing resources and support (ISCRA-C ``ARTTESC''), and NVIDIA Corporation for the donation of the Titan X Pascal GPU. M. B. and P. N. acknowledge financial support from Slovenian Research and Innovation Agency through grants J2-2502, L2-4455 and J2-50092.
F.M. is kindly acknowledged for fruitful discussion.

\end{document}